# GENETIC CODE: A NEW UNDERSTANDING OF CODON – AMINO ACID ASSIGNMENT

Zvonimir M. Damjanović* and Miloje M. Rakočević**

*Montenegrin Academy of science and arts (CANU), Podgorica, Montenegro*
*** Department of Chemistry, Faculty of Science, University of Niš, Serbia*
(E-mail: m.m.r@eunet.yu)

**Abstract**.

In this work it is shown that 20 canonical amino acids (AAs) within genetic code appear to be a whole system with strict AAs positions; more exactly, with AAs ordinal number in three variants; first variant 00-19, second 00-21 and third 00-20. The ordinal number follows from the positions of belonging codons, i.e. their digrams (or "doublets"). The reading itself is a reading in quaternary numbering system if four bases possess the values within a specific logical square: $A = 0, C = 1, G = 2, U = 3$. By this, all splittings, distinctions and classifications of AAs appear to be in accordance to atom and nucleon number balance as well as to the other physico-chemical properties, such as hydrophobicity and polarity.

K e y  w o r d s: Genetic code, Genetic code table, Translation, Numbering system, Spiral model of Genetic code, Canonical amino acids, Logical square, Hydrophobicity, Polarity, Hydropathy, Perfect numbers, Friendly numbers.

## 1 INTRODUCTION

Gamow was the first who attempted to resolve the problem of codon – amino acid assignment, i.e. to answer the question how 64 codons can have the possible meanings for 20 canonical amino acids (AAs). His solution was the so called "diamond code" (Gamow, 1954) in which 64 codons were classified into 20 classes, corresponding to 20 AAs (Hayes, 1998: "Symmetries of the diamond code sort the 64 codons into 20 classes ... All the codons in each class specified the same amino acid"; cf. legends to Figures 2 and 5 in cited paper). Unfortunately, the real genetic code (Crick, 1966, 1968; Patel, 2005) appears to be less regular than Gamow's. Indeed, the experimentally observed results showed that the codon – amino acid assignment realizes through the relationships very different of those, postulated by the diamond code (Tables 1 and 2 in relation to Table 3).

| Amino acid | R-group property | Mol. weight | Class | Secondary propensity |
|---|---|---|---|---|
| Gly | Non-polar | 75 | II | turn |
| Ala | aliphatic | 89 | II | α |
| Pro |  | 115 | II | turn |
| Val |  | 117 | I | β |
| Leu |  | 131 | I | α |
| Ile |  | 131 | I | β |
| Ser | Polar | 105 | II | turn |
| Thr | uncharged | 119 | II | β |
| Asn |  | 132 | II | turn |
| Cys |  | 121 | I | β |
| Met |  | 149 | I | α |
| Gln |  | 146 | I | α |
| Asp | Negative | 133 | II | turn |
| Glu | charge | 147 | I | α |
| Lys | Positive | 146 | II | α |
| Arg | charge | 174 | I | α |
| His | Ring/ | 155 | II | α |
| Phe | aromatic | 165 | II | β |
| Tyr |  | 181 | I | β |
| Trp |  | 204 | I | β |

**Table 1.** Amino acid properties. This Table is downloaded from Table 1 in Patel (2005): "Properties of amino acids depend on their side chain R-groups. Larger molecular weights indicate bigger side chains. The 20 naturally occurring amino acids divided into two classes of 10 each, depending on the properties of aminoacyl-tRNA synthetases that bind the amino acids to tRNA. The dominant properties of amino acids for forming secondary protein structures are also listed".

At the same time when Crick brought out two possible hypotheses on the codon – amino acid assignment (genetic code was frozen in an evolution process on the level "64 codons : 20 AAs", or this ratio is result of the stereochemical conditions), a special understanding came from Y. Rumer (Rumer, 1966; Konopel'chenko & Rumer, 1975) [Rumer, 1966, p. 1393: "Considering the group of codons, that relates to one and the same amino acid, shows that within every codon (z | yx) (it should be read from right to left side) it is expedient to separate two-letter 'root' | yx) of the 'end' (z |. So, every amino acid, in a general case, has a corresponding and specific root, and degeneration of the code appears as consequence of exchanging of the endings."]. Unfortunately, this understanding was forgotten two next decades, when it was restored by Shcherbak still once (Shcherbak, 1989, 1993, 1994).

Table 2
Mitochondrial genetic code

| | | | |
|---|---|---|---|
| **UUU** Phe | UCU Ser | UAU Tyr | UGU Cys |
| **UUC** Phe | UCC Ser | UAC Tyr | UGC Cys |
| UUA Leu | UCA Ser | UAA Stop | UGA Trp |
| UUG Leu | UCG Ser | UAG Stop | UGG Trp |
| | | | |
| CUU Leu | CCU Pro | CAU His | CGU Arg |
| CUC Leu | CCC Pro | CAC His | CGC Arg |
| CUA Leu | CCA Pro | CAA Gln | CGA Arg |
| CUG Leu | CCG Pro | CAG Gln | CGG Arg |
| | | | |
| AUU Ile | ACU Thr | AAU Asn | AGU Ser |
| AUC Ile | ACC Thr | AAC Asn | AGC Ser |
| AUA Met | ACA Thr | AAA Lys | AGA Stop |
| AUG Met | ACG Thr | AAG Lys | AGG Stop |
| | | | |
| GUU Val | GCU Ala | GAU Asp | GGU Gly |
| GUC Val | GCC Ala | GAC Asp | GGC Gly |
| GUA Val | GCA Ala | GAA Glu | GGA Gly |
| GUG Val | GCG Ala | GAG Glu | GGG Gly |

**Table 2**. Mitochondrial genetic code. This Table is downloaded from Table 2 in Patel (2005): "The (vertebrate) mitochondrial genetic code differs slightly from the universal genetic code. The wobble rules are exact for the mitochondrial code, so the third codon position has only a binary meaning. Class II amino acids are indicated by boldface letters". (Note: The differences in standard genetic code: AUA – Ile, UGA – Stop, AGA – Arg, AGG – Arg).

## 2 PRELIMINARIES

Our today's understanding of codon–amino acid assignment relies on the said Rumer's conception; on the other words, our understanding follows, in principle, from a specific manner of reading the codons and their digrams, i.e. „doublets", in which AAs are given in the strict ordinal numbers through three variants (presented in this paper in quaternary and/or in decimal numbering system).

**First variant** with ordinal number of AAs **00-19** (cf. Damjanović, 1998, and Appendix 1, Surveys 1 & 2 in this paper): 00 K; 01 Q; 02 E; 03 ⊗; 04 T; 05 P; 06 A; 07 S; 08 S, R; 09 R; 10 G; 11 C, W, ⊗; 12 I, I[1], M; 13 L; 14 V; 15 F, L; 16 N; 17 H; 18 D; 19 Y; **Second variant** with ordinal number of AAs **00-21** (cf. Solutions 1- 4 in Section 3, Figure 1 and Tables 4 and 5): 00 K; 01Q; 02 E; 03 ⊗; 04 T; 05 P; 06 A; 07 S; 08 ∅; 09 R; 10 G; 11 C; 12 I; 13 L; 14 V, 15 F; 16 N; 17 H; 18 D; 19 Y; 20 W, 21 M; **Third variant** with ordinal number of AAs **00-20** (cf. Figure 2 and Tables 3, 6, 7 and 8): 00 K, 01Q; 02 E; 03 ⊗; 04 T; 05 P; 06 A; 07 S; 08 ∅ ; 09 R; 10 G; 11 C; 12 I; 13 L; 14 V, 15 F; 16 N; 17 H; 18 D; 19 Y, W, 20 M[2].

This way, it is presented a logic from which a series of AAs 00-19 follows, with the interruption of ordinal number 3 (for all three "stop" codons as a "stop" command within AAs alphabet), and also the logic from which follows a series of AAs 00-21, with interruption of ordinal numbers 3 and 8 (3 as a "stop" command and 8 as a "phantomic" interruption, an empty space) (Damjanović, 1998; Rakočević, 2004). In the other words, in the first case of reading, a "two-meaning" logical pattern is presented while in the second case a "one-meaning" logical pattern. For example, the serine is located on two locations (7 and 8) in the "two-meaning" logical pattern and only on one location (7) in the "one-meaning" logical pattern. Accordingly, for arginine, in the second case, we must assume that it is located only in the position 9, since otherwise it would be "mixed" with serine on the position 8. On the other hand, since position 11 must be occupied with priority by cysteine, then tryptophan must be moved for one cycle, according to the module 9, and it should appear on the position 20 (11+ 9 = 20) as a neighbor of tyrosine on the position 19 (neighbors in GCT also through a "stop command" loop, encoding by three "stop" codons). Similar case happens with methionine which, in relation to isoleucine, moves for one modular cycle further, on the position 21 (12+9 = 21) as a "neighbor" of tryptophan;

---

[1] In Shcherbak's four-codon/non-four-codon AAs system (Fig. 1 in Shcherbak, 1994) there are only three AAs as "duplicates" (L, S and R), whereas here appears isoleucine as a fourth (through Pu / Py coding codons).

[2] Within decimal numbering system (q = 10), the triplet 19-20-21 appears to be very adequate from still one very specific manner. Namely, if we exclude the zeroth amino acid (K), then the sum of ordinal numbers of other 19 AAs equals 190 – 10 units pro each one of amino acids. On the other hand, the sum from 0 to 20 (20 AAs plus one "stop' command) equals 210 – 10 units pro each one entity, amino acid or "stop' command. [Cf. the determination through "the symmetry in the simplest case" (Marcus, 1989), through the pair (q= 10 and q/2 = 5), presented in legend of Table 8].

"neighbor", from the aspect of the existence of one-meaning, i.e. one-codon amino acids. In the third case of reading (Damjanović and Rakočević, 2005) we have the appearance of a specific "mobile loop". Regarding Figure 2 we see that tryptophan comes one step back "in order" to be together with tyrosine (cf. legend of Table 3) and, at the same time, methionine comes at the former position of tryptophan. This "mobile loop" follows from a "theory of ribosomal code" (Section 4), from amino acid positions in Genetic Code Table (GCT), and from physico-chemical properties of AAs. [W and M as only two one-codon AAs; M and T as horizontal neighbors in GCT and only two AAs (within alanine stereochemical type) with a $CH_3$ atom group, etc.].

The codons for W and M in Table 3 are given in a vice versa position in order to signalize the mobile loop valid for the codons and not only for AAs. (last column: O.Nr.): The amino acid Y is $103_4$ in a normal reading: UAC/CAU → $103_4$; the M possesses the same ordinal number, reading from anticodon: AUG/GUA → CAU → $103_4$; on the other hand, the M possesses the ordinal number $111_4$, reading from the syn-codon: AUG/CCC → $111_4$. The arithmetical mean for M equals $103_4 + 111_4 = 110_4$. which ordinal number possesses also the W, reading from its anticodon: UGG/GGU → CCA → $110_4$. Thus, methionine – the first amino acid in protein biosynthesis – possess all three meanings ($103_4$, $110_4$. and $111_4$), valid for the whole AAs system.

## 3  A HOLISTIC APPROACH

For a better understanding the process of codon – amino acid assignment, a holistic approach to genetic code is needed (Rakočević, 1998, 2005). In such an approach genetic code appears to be a harmonic structure – a whole (and full) system, determined by Golden mean as well as by Generalized Golden Mean [GGM: three equations $(x^2 \pm 1 - 1 = 0; x^n + x - 1 = 0; x^2 + x - m/2 = 0; n = 1, 2, 3, ...; m = 0, 1, 2, 3 ...)$(cf. math.GM/0611095) in correspondence to positions of 20 canonical AAs and 64 codons on the segment-line 0-63, within a binary-code tree]; also as the first and only one possible case from many aspects. So, within a system of "letter-root-word-alphabet", the first possible case is the system "1-2-3-4" (64 *three-letter* words, changeable exactly for *one* letter accordingly to Gray code, 16 *digrams* and *four-letter* alphabet). From the aspect of validity of principle of minimum change and principle of continuity, any other case is not possible.

The same follows from the aspect of information theory, accordingly to principle of symmetry and the principle of self-similarity, in relation to the real three-dimensionality. Namely, only a 6-bit binary-code tree with 64 words within a $B^6$ Boolean hyper-cube is possible (Rakočević, 1998).

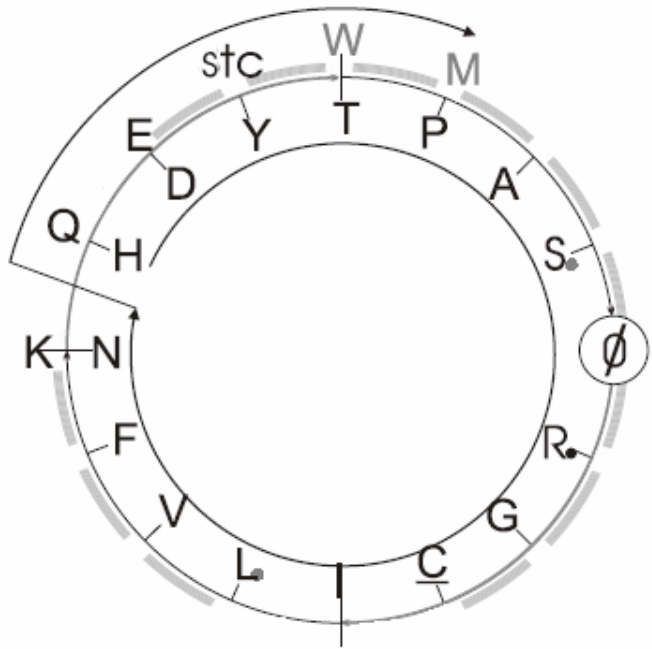

**Figure 1**. Spiral model of Genetic code (I). The spiral order of amino acids in accordance to their ordinal number, given by first and second variant of reading from codons and/or their digrams (doublets); the reading in quaternary numbering system, as it is presented in Remark 1 on p. 13 (first variant: exactly in relation to App. 1, Survey 2; second variant: exactly in relation to Solutions 1 – 4 and Table 4). An analogous figure for third variant is Figure 2; while a cross arrangement of this spiral is given in Figures 3 and 4.

[Distribution on a 7-bit tree: ($B^7$- $B^6$- $B^5$- $B^4$- $B^3$- $B^2$- $B^1$- $B^0$ / 1 x 128, 2 x 64, 4 x 32, **8 x 16**, 16 x 8, 32 x 4, 64 x 2, 128 x 1); on a 6-bit tree: ($B^6$- $B^5$- $B^4$- $B^3$- $B^2$- $B^1$- $B^0$ / 1 x 64, 2 x 32, 4 x 16, **8 x 8**, 16 x 4, 32 x 2, 64 x 1); on a 5-bit tree: ($B^5$- $B^4$- $B^3$- $B^2$- $B^1$- $B^0$ / 1 x 32, 2 x 16, **4 x 8**, 8 x 4, 16 x 2, 32 x 1)]. By this, the principle of self-similarity ($B^3$ Boolean real cube, expressed through a 3D model, and/or only through a holographic model, with **8** classes, each class with **8** words i.e. codons) is related to a determination by perfect and friendly numbers as follows (about perfect and friendly numbers as determinants of genetic code, see in Rakočević, 1997 and www.sponce.net).

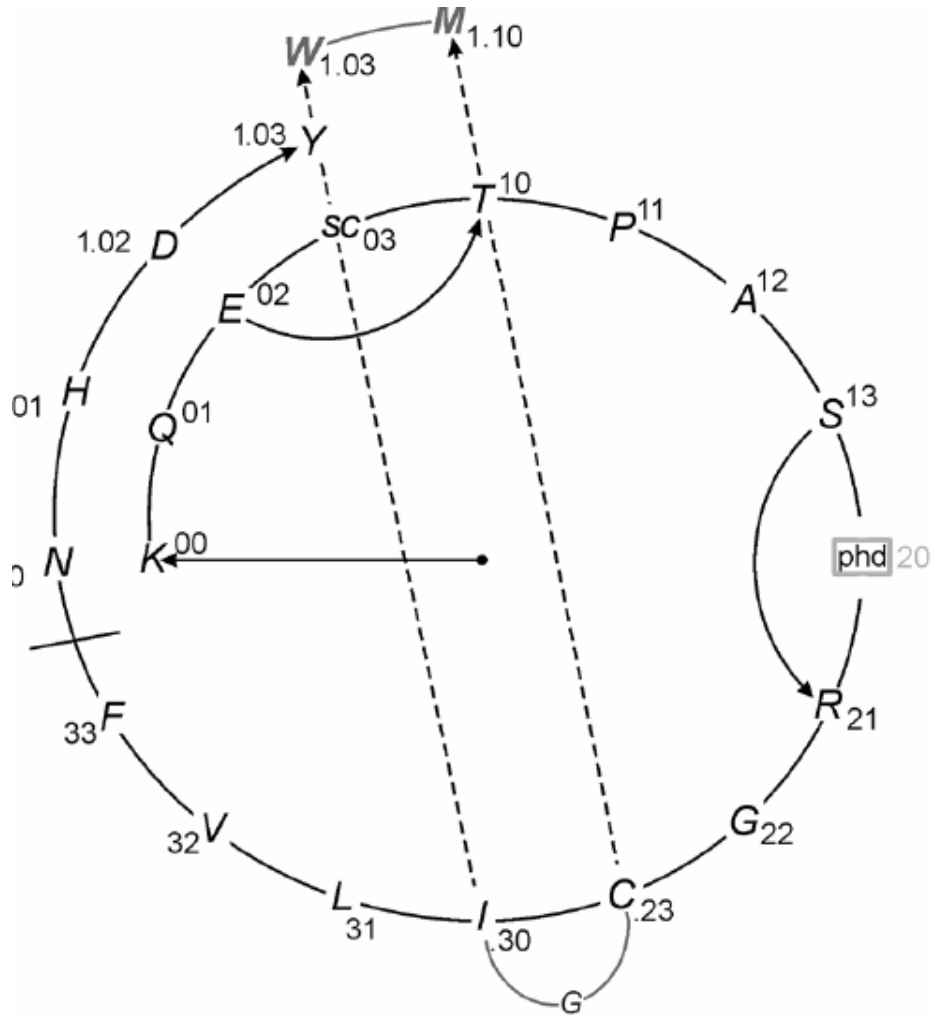

**Figure 2**. Spiral model of Genetic code (II). The spiral order of amino acids in accordance to their ordinal number, given by third variant of reading from codons and/or their digrams (doublets); the reading in quaternary numbering system, as it is presented in Remark 1 on p. 13; the arrangement itself exactly in relation to Tables 3, 6 and 8. A cross arrangement of this spiral is given in Figures 5.

| m/codon | rs/codon | c/ama | O.Nr. |
|---|---|---|---|
| pu/AA | /00 | K | 0 |
| pu/AC | /01 | Q | 1 |
| pu/AG | /02 | E | 2 |
| .yx | /yx | T | 4 |
| | | P | 5 |
| | | A | 6 |
| | | S | 7 |
| py/GC | /21 | R | 9 |
| py/GG | /22 | G | 10 |
| py/GU | /23 | C | 11 |
| notG/UA | /30 | I | 12 |
| .yx | /yx | L | 13 |
| | | V | 14 |
| py/UU | /33 | F | 15 |
| py/AA | 1/00 | N | 16 |
| py/AC | 1/01 | H | 17 |
| py/AG | 1/02 | D | 18 |
| py/AU | 1/03 | Y | 19 |
| G/UA | 1/03 | W | 19 |
| G/GU | 1/10 | M | 20 |
| 00 | K, N | | |
| 01 | Q, H | | 77 |
| 02 | E, D | | |
| 03 | ⊗, Y | | |
| | | | |
| 10 | W, T | | |
| 11 | M, P | | |
| 12 | A | 54 | |
| 13 | S | | |
| | | | 77 |
| 20 | Ø | | |
| 21 | R | | |
| 22 | G | 23 | |
| 23 | C | | |
| | | | |
| 30 | I | | |
| 31 | L | | |
| 32 | V | | 50 |
| 33 | F | | |
| | (N) | | |

**Table 3 (left)**. The relations among codons and canonical amino acids. The "m/codon" designates messenger RNA codons. In relation to Genetic Code Table (GCT) all codons are read from right to left; "pu"/"py" – purine or pyrimidines in third codon position; the point within codon "**.**yx" designates that both purine or pyrimidines can be in third codon position; "notG" means that only guanine is not possible in third position of codons, coding for amino acid isoleucine; The "rs/codon" – ribosomal codons and/or the conditions within ribosome system, enable to read the ordinal numbers in quaternary numbering system; the digram, i.e. doublet "/yx" is an analog of digram "**.**yx" in column "m/codon" (/yx equals $31_4$ for leucine and $32_4$ for valine). The "c/ama" – canonical amino acids; at the end of this column there are three multi-meaning AAs: Y, W and M, multi-meaning in terms of ordinal numbers. Last column: O.Nr. – Ordinal number.

**Table 4 (right)**. Ordinal number of amino acids in relations to codon digrams. The splitting (classification and/or grouping) of canonical amino acids in accordance to their ordinal number, given by Figure 1, i.e. by second variant of spiral model of genetic code, as it is presented in Remark 1 on p. 13, and to logical square (0-1-2-3) at the same time. An analogous table for third variant is Table 6, while (table building) for first variant is leaving to the readers (by this one must consult Surveys 1 and 2 in Appendix 1). A "mirror image" table (analogous to Table 7) also is leaving to the readers. Atom number balance, presented here, and valid for three AAs groups (50-77-77) stay in correspondence with Golden mean balance (60-66-78) as it is shown in Survey 3 in Appendix 2; in correspondence by principle of minimum change as well as of continuity, through an evident symmetry.

| 00 | K, N |     |      | 00 | K, N |    |
| 11 | M, P |     |      | 01 | Q, H |    |
| 22 | G    | 57  |      | 02 | E, D | 95 |
| 33 | F    |     |      | 03 | Y, W |    |
|    |      | (11)|      |    |      |    |
| 01 | Q, H |     |      | 10 | T, M |    |
| 02 | E, D |     |      | 11 | P    |    |
| 03 | ⊗, Y | 68  |      | 12 | A    | 36 |
| 12 | A    |     |      | 13 | S    |    |
| 13 | S    |     |      |    |      | 59 |
| 23 | C    |     |      | 20 | Ø    |    |
|    |      | (11)|      | 21 | R    |    |
| 10 | W, T |     |      | 22 | G    | 23 |
| 20 | Ø    |     |      | 23 | C    |    |
| 30 | I    | 79  |      |    |      |    |
| 21 | R    |     |      | 30 | I    |    |
| 31 | L    |     |      | 31 | L    |    |
| 32 | V    |     |      | 32 | V    | 50 |
|    |      |     |      | 33 | F    |    |

**Table 5 (left).** Ordinal number of amino acids in relations to codon digrams. The splitting (classification and/or grouping) of canonical amino acids in accordance to their ordinal number, given by Figure 1, i.e. by second variant of spiral model of genetic code, as it is presented in Remark 1 on p. 13, and to logical square (0-1-2-3) at the same time. An analogous table for third variant is Table 6, while (table building) for first variant is leaving to the readers (by this one must consult Surveys 1 and 2 in Appendix 1). A "mirror image" table (analogous to Table 7) also is leaving to the readers. Atom number balance, presented here, and valid for three AAs groups (50-77-77) stay in correspondence with Golden mean balance (60-66-78) as it is shown in Survey 3 in Appendix 2; in correspondence by principle of minimum change as well as of continuity, through an evident symmetry.

**Table 6 (right)**. The relations among canonical amino acids (I).The splitting (classification and/or grouping) of canonical amino acids in accordance to their ordinal numbers, given by Figure 2, i.e. by third variant of spiral model of genetic code, as it is presented in Remark 1 on p.

13, and to logical square (0-1-2-3) at the same time. An analogous table for second variant is Table 4. The atom number balance, presented here, and valid for three AAs groups (95-59-50) is explained in the text (Section 4).

| | | | | | | |
|---|---|---|---|---|---|---|
| 00 | K, N | | | 000 | K | |
| 10 | T, M | | | 001 | Q | 36(06) |
| 20 | *G* | 56-1 | | 002 | E | |
| 30 | I | | | 003 | * | |
| | | | | | | 61(28) |
| | | | | 010 | T | |
| 01 | Q, H | | | 011 | P | 25(22) |
| 11 | **P** | | | 012 | A | |
| 21 | **R** | 60 | | 013 | S | |
| 31 | L | | | 020 | *G* | |
| | | | | 021 | R | |
| 02 | E, D | 56+36 | | 022 | G | 23(30) |
| 12 | **A** | | | 023 | C | |
| 22 | **G** | 60-28 | | 030 | I | 73(84) |
| 32 | V | | | 031 | L | |
| | | | | 032 | V | 50(54) |
| | | | | 033 | F | |
| 03 | Y, W | | | 100 | N | |
| 13 | **S** | | | 101 | H | 70(90) |
| 23 | **C** | 56+1 | | 102 | D | |
| 33 | F | | | 103 | Y,W | |
| | | | | 110 | M | |

**Table 7 (left)**. The relations among canonical amino acids (II). This Table is a "mirror image" of Table 6 {[(00-01-02-03) (10-11-12-13) (20-21-22-23) (30-31-32-33)] / [(00-10-20-30) (01-11-21-31) (02-12-22-32) (03-13-23-33)]}. The atom number balance appears to be in correspondence with first and second perfect number ($56 = 2 \times 28$; $36 = 6^2$; $60 = 6 \times 10$ etc.).

**Table 8 (right)**. The splitting of amino acids in correspondence to hydrophobicity. The splitting (classification and/or grouping) of canonical amino acids in accordance to their ordinal number, given by Figure 2, i.e. by third variant of spiral model of genetic code, as it is presented in Remark 1 on p. 13. An analogous table for first and second variant is leaving to the readers. Atom number balance, presented here, and valid for three AAs groups (*70-61-73*) stay in correspondence with Golden mean balance by an evident symmetry ["the symmetry in the simplest case" (Marcus, 1989), through the base of decimal numbering system, q = 10; q/2 = 5 (*70*-60 = **10**; 66-*61* = **5**; 78 – *73* = **5**). The splitting of AAs corresponds also with hydrophobicity and polarity of AAs (see the text: Section 4). Bold amino acids as in Figure 5.

The sum of the ordinal numbers on two middle and neighbor branches on the 6-bit binary-code tree is 220 and 284, respectively, which two numbers represent first pair of friendly numbers. Their sum equals 504, as in two second, two third and two fourth branches (reading one branch from left and second branch from right side at the same time, in relation to middle point of tree). On the other hand, we have the realization each of four possible letters, minimum once, through the realization of first four words (0. UU**U**, 1. UU**C**, 2. UU**A**, 3. UU**G**) on the binary-code tree and/or in GCT (cf. Rumer's and Shcherbak's idea about the four-codon AAs in next Section). The sum of their ordinal numbers equals 6, which is the first perfect number. After the realization of first eight words (0. UU**U**, 1. UU**C**, 2. UU**A**, 3. UU**G,** 4. **C**U**U**, 5. **C**U**C**, 6. **C**U**A**, 7. **C**U**G**) occurred the determination of upper half of GCT (4 half-columns, each half-column with 8 codons), and the sum of all eight ordinal number equals 28, which is the second perfect number. Stepping to the ordinal (codon) number 31, we have the determination of the left half of GCT (2 columns, each column with 16 codons) and the realization of third perfect number, because the sum of all numbers from 0, i.e. from 1 to 31 equals 496. Finely, with a full cycle (from 0 to 63 and back, from 64 to 127) we have the determination of the full GCT system and the realization of fourth perfect number at the same time (8128), because the sum from 1 to 127 equals 8128.

The holistic approach to genetic code comes also from some specific chemical aspects. First of all, from the aspect of classification of 20 canonical AAs into four stereochemical types (Popov, 1989; Rakočević and Jokić, 1996) – glycine type (with only G amino acid), proline type (only P), valine type (V & I) and alanine type (the rest of 16 AAs, each amino acid with a H-C-H "screen" between the "head" and "body", i.e. side chain; the exception is threonine with an H-C-CH$_3$ "screen"). The appearance of glycine corresponds to the appearance of the first possible non-hydrocarbonity (H as the side chain); the appearance of alanine corresponds to the appearance of the first possible non-cyclic hydrocarbonity (CH$_3$ as side chain); the appearance of valine corresponds to the appearance of the first possible hydrocarbon half-cyclity (isopropyl atom group as side chain), and the appearance of proline corresponds to the appearance of the first possible hydrocarbon cyclicity (–CH$_2$–CH$_2$–CH$_2$– group in side chain and in contact with the "head").

As a second holistic chemical aspect is the splitting of 20 canonical AAs into two classes, 10+10, in correspondence to two classes of enzymes aminoacyl-tRNA synthetases. The simpler and/or smaller AAs (within AAs pairs) are handled by less complex enzymes of class II, whereas larger (more complex) AAs molecules are handled by more complex enzymes of class I (bold underlined) through a pairing process: I. aliphatic AAs – Ia. hydrocarbon non-polar AAs (G-**V**, P-**I**, A-**L**); Ib. chalcogene polar AAs (S-**C**, T-**M**, N-**Q**); Ic. polar charged AAs (D-**E**, K-**R**); II. aromatic AAs (F-**Y**, H-**W**). [Cf. this classification with a similar but more global, presented here in Table 1, in relation to Table 2 (Patel, 2005, p. 529: "A closer inspection of Table 2 shows that all class-II amino acids, except Lys, can be coded by the codons NNY") (Y = U or C)]. A further classification is also a proof for the wholeness within a holistic system of genetic code, the classification in relation to the base type in third position of the belonging codon. Thus, in class II there are AAs whose codons do not possess purine in third position (first subclass): N, D, F, H (neighbors within the system presented in Figures 1 and 2 as F, N, H, D) with 40 atoms within their side chains; then AAs whose codons possess purine in third position (second subclass): **K**, P, A, S, T, G (also as neighbors in Figures 1 and 2, as T, P, A, S, excluding K, which is an exception with the location on the other side in Figures 1 and 2; the exeption still once as in above given Patel's comment; also as a zeroth amino acid in three variants of reading within AAs system, presented in first paragraph of Section 2) with 40+01 atoms. On the other hand in class I there are AAs whose codons possess either pyrimidine or purine in third codon position (first subclass): V, L, R, also with 40 atoms within their side chains; then AAs whose codons possess only purine in third codon position (second subclass): M, Q, E, W with 40+10 atoms; as a third subclass there are AAs whose codons possess only pyrimidine in third codon position: C & Y with 40-20 atoms within their side chains. Out of the classification within class I there is isoleucine which belongs to the first subclass within standard genetic code and to the third subclass within mitochondrial genetic code, presented in Table 2 (cf. isoleucine positions in Tables 3-5 and in App. 1, Surveys 1 and 2).

The presented holistic approach can be also interesting in the study of possible analogy with other natural codes, esspecialy with visual code. In shortest words, the genetic code alphabet UCAG can be analogue with visual alphabet UBGR (U-Union of all rainbow colors, i.e "white" color (light); B – Blue; G – Green, and R – Red). Namely, as it is known, each human cone cell absorbs light in only one of three bands of the spectrum: blue, green and red. This follows from the fact that there exist three types of the genes, coded for the three color receptor proteins. [Note: Cone cell is one of specialized, photosensitive cells in the retina of the eye concerned with the perception of color and with daylight vision]. Within 64 possible "words" from the alphabet UBGR there are exactly 28 (second perfect number!) words which possess U, as white light, and $6^2 = 36$ "color words" without U (number 6 as first perfect number). The hypothesis about a possible analogy of genetic code and visual code is only a part of a larger idea on the analogy of genetic code and neuro code, and separately – genetic code and a sensory code (Damjanović, 1998).

## 4 DIVERZITY OF CODON – AMINO ACID ASSIGNMENT

Our earlier studies (Damjanović, 1998; Rakočević, 2002, 2004; Damjanović & Rakočević, 2005) of codon – amino acid assignment have lead us to a specific numbering coding of nucleotides in correspondence with a logical square (A = 0, C =1, G =2, U =3)[3], and to a possible their reading through quaternary numbering system; also to an ordering of digrams (base doublets within the codon) and codons (reading right-to-left: **.**yx & Z**.**yx) as ordinal numbers from $000_4$ to $111_4$ for belonging canonical AAs. In this way, sixteen digrams and belonging AAs appear as a parallel discrete array (the nucleotide-letters within the codons must be read from right-to-left and the numbers, in quaternary numbering system, from left-to-right): *AA*(00) – K, N; *AC* (01) – Q, H; *AG*(02) – E, D; *AU*(03) – stc,Y; *CA*(10) – T; *CC*(11) – P; *CG* (12) – A; *CU*(13) – S; *GA*(20) – S, R; *GC*(21) – R; *GG*(22) – G; *GU*(23) –C, W, stc; *UA*(30) – I, I, M; *UC*(31) – L; *UG*(32) – V; *UU*(33) – F, L; The ordinal numbers from $100_4$ to $103_4$ must be read from respective codons as follows (Tables 3, 4 and 5): *C.AA*(1.00) – N; *C.AC*(1.01) – H; *C.AG*(1.02) – D; *C.AU* (1.03) – Y.

(*Remark* 1: We use here the point in designations accordingly to the convention, given obove as „**.**yx" & „Z**.**yx". By this one must notice that at last four AAs the codon letter "C" has a meaning of "Z" in „Z**.**yx" ).

The digram *GA*, staying for both S & R, appears as an "phantomic" digram, because these two AAs possess their "first" digrams from the codon families: *CU* and *GC*, respectively. Otherwise said, here there are two possibilities for reading. As a first we have a "two-meaning" logical patterns, where both serine and arginine are located on two locations each [*CU*(13) – S; *GA*(20) – S, i.e. 7 and 8 in decimal numbering system; *GA*(20) – R; *GC*(21) – R, i.e. 8 and 9 in decimal numbering system]. In the second case – an "one-meaning" logical patterns, where both serine and arginine are located on one location each, and between them appears a "phantomic" digram (phd) with an empty space for non-existing amino acid [*CU*(13) – S; *GA*(20) ∅, *GC*(21) – R].

---

[3] As to now we founded only one paper with the same numbering coding for bases, i.e. nucleotides; however, not for RNA but for DNA: A → 0, C → 1, G → 2, T → 3 (Sirakoulis et al, 2004).

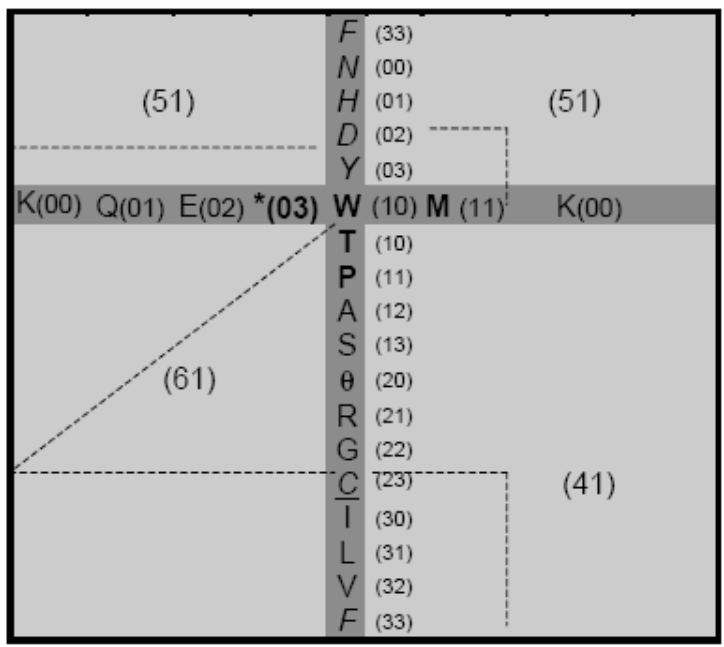

**Figure 3.** Cross model of Genetic code (I). The cross arrangement of amino acids displayed from Figure 1 in accordance to an atom number balance: above/down 102/102 of atoms within amino acids side chains.

In such an understanding we find that serine must be located only on one location (position 7) and arginine only in one position (the position 9), since otherwise these two AAs would be "mixed" on the position 8. From this reason it is clear why position 8 must be empty (designation ∅ in corresponding Figures and Tables). Ordinal numbers for W and M and their respective analogs deserve a particular consideration: (I). W, T → 10 and/or 1.10 and M, P → 11 and/or 1.11, in Figures 1, 3 and 4 and in Tables 4 and 5; (II). W, Y → 03 and/or 1.03 and M, T → 10 and/or 1.10, plus P → 11 in Figures 2 and 5, and in Tables 3, 6, 7 and 8. In order for a better understanding, the first case is displaying in Solutions 1 – 4 still once:

**K(00)-Q(01)-E(02)-\*(03)**-T(10)-P(11)-A(12)-S(13)- ∅ **(20)**
 R(21)-G(22)-C(23)-. I(30)-L(31)-V(32)-*F(33)*            (1)

*F-N(00)-H(01)-D(02)-Y(03)-W(10)-M(11)*                   (2)
 *(100)- (101)-(102)-(103)-   (110)-(111)*

**K(00)-Q(01)-E(02)-*(03)**- W(10)-M(11)-A(12)-S(13)- Ø **(20)**  
  R(21)-G(22)-<u>C</u>(23)- `. I(30)-L(31)-V(32)-*F(33)*         (3)

*F-N(00)-H(01)-D(02)-Y(03)*- **T(10)-P(11)**                     (4)  
  *(100)- (101)- (102)-(103)-* (110)- (111)

**Figure 4**. Cross model of Genetic code (II). The cross arrangement of amino acids displayed from Figure 1; the same as in Figure 3 except a vice versa position for the pairs W-M / T-P. The atom number balance follows a molecule number balance. [Notice that in a vice versa position for P/M we have the situation 1(11) – 1 molecule with 11 atoms].

    As we can see from Solutions (1) and (2) the spiral model of genetic code, relating to Siemion and Siemion's rule, as well as Davidov's rule (Siemion and Siemion, 1994; Davidov, 1998; cf. Rakočević, 2004)[4] can be given in form of "a

---

[4] Classification of canonical AAs derived from our dynamic model brings about clarification of physicochemical criteria, such as purity, pyrimidinity – and, particularly, codon rules. The system implies both rules of Siemion and Siemion and of Davidov, as well as balances of atom and nucleon numbers within groups of AAs. Formalization in this way opens fruitful chances of extrapolating backwards, to initial organization of heredity.

cross" too. [The Solutions (3) and (4) are the same as Solutions (1) and (2), respectively, with a position changing for two AAs pairs: T-P/ W-M]. Bearing in mind that both variants – spiral and cross – seek a connection "head to tile" (F-F), two intersecting lines appear (cf. Figures 1, 3 & 4 with the above given Solutions); the horizontal (shorter) leg of the cross consists of AAs of Pu type (K, Q, E, W, M), while the vertical (longer) leg contains two sub-classes: up there are AAs of Py type (italic: F, N, H, D, Y), and down there are AAs of "Py or Pu" type (from T to V-F), with an exeption of cysteine which is of Py type.

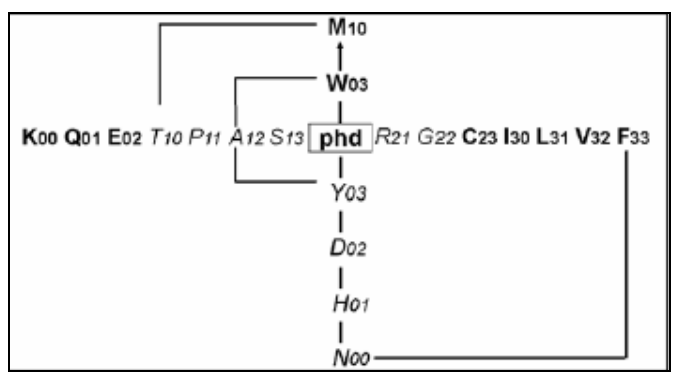

**Figure 5**. Cross model of Genetic code (III). The cross arrangement of amino acids displayed from Figure 2 in accordance to physico-chemical parameters, hydrophobicity and polarity (explanation in the text: Section 4). Bold amino acids as in Table 8.

### 4.1. The reading of codons and their digrams

The reading of codons and their digrams in quaternary and/or decimal numbering system (according to Damjanović, 1998) starts with "zero" column in GCT where are the codons with middle base "A". Accordingly, we read zeroth "digram" AA as 00 within the codon AAA that is coding for lysine. Subsequently, the neighboring codon AAC, in reverse case as CAA (that is as $100_4$), which is the number $16_{10}$, as ordinal number for asparagine, etc. In such a manner the 18 AAs can be read from the codons; the first to last, the tryptophan, is read from the anticodon (anticodon ACC, read from right to left as CCA) with ordinal number $110_4$, that is $20_{10}$ (Damjanović, 1998, p. 6: "the cycle of digrams is presented, and the spiral of codons ... with the 'inverse' appearance of number 20"). Despite the last amino acid, methionine, can be read in a specific way from the "ribosomal code", with ordinal number $111_4$, that is $21_{10}$, it is interesting that ordinal number of methionine can be read also from the "syn-codon" (Rakočević, 2004, p. 226 down).

[*Remark* 2. The codons with the largest diversity (all three bases are different) together with codons of the smallest diversity (all three bases are the same) contain a balanced number of nucleons: within side chains of AAs assigned to 12 codons with a clockwise direction (U → C→ A) plus codons UUU & AAA there are 703 nucleons; the same result is in side chains of AAs assigned to 12 codons with an anti-clockwise direction plus the codons CCC & GGG (cf. Fig. 4 in Rakočević, 2004, p. 226). From that it is a reason for introduction of the term "syn-codon" on the following way: UUU is a syn-codon for all six permutations of codon CAG, with the designation UUU/CAG. The same goes for the remaining three syn-codons: CCC/UAG (where one of the permutations is methionine-codon AUG), than AAA/UCG and, finally, GGG/UCA].

**4.2. Theory of a ribosomal code**

As an extract from translation, it is possible a specific mapping of codons onto canonical amino acids (m/codons onto c/ama in Table 3). By this all "capital" and "stop" codons, that impact on ribosome assembling, as well as the duplicates of canonical amino acids (S, R, L and I), are left aside from the model, presented in Fig. 2 & 5). Modeling itself runs, as we said, with help of quaternary numeric transforms of 4 nucleotides: A=0, C=1, G=2, U=3. Codons **xy/Z,** if read from right, expose 16 digrams •**yx** (cf. Remark 1 on p. 13) as a smooth array of ordinal numbers (O.Nr. in Table 3), where 4 groups of digrams, dominated by A, C, G and U in central **y** codon position represent *complemental* pairs **Ax-Ux** and **Cx-Gx** (Table 3) making a quasi cycle, mapped onto canonical amino acids from K to **F** (Fig. 2).

The fact that ribosome is ubiquitous medium of convergence of 61 messenger RNA codons to 20 crucial tRNA canonical AAs blocs is reflected in a *Theory of ribosomal coding* which includes the following principles:

 (a) Numeric values of **Z** are restricted to Ø and 1, corresponding to A and C. This, for example, allows interpretation of "ribosomal" codons mapped onto **N, H, D** and **Y** (cf. Remark 1 on p. 13) through  possible "para-codons" (De Duve, 1988).

(b) Sliding of ribosome along mRNA is discrete, elementary step covering 3 nucleotides (i.e. codon); the adherence of **y** to its complement (**y**|cy) makes coupling center of codon to tRNA.

(c) The event of translation happens within polar space of ribosome. So the numeric code includes space angles: as mRNA is, by no means, a straight line, a linear arrangement is approximated by a series of **y**|cy, which is the center of polar space.

(d) Translation is made readable with help of an abstract of "ribosomal" codon ("rs" codon) as well as rs**Z** (Ø: 0 or absent; 1: C engaged as **Z** in third codon position; C, writing in "cy" as a small letter, instead a large letter because "C", i.e. "c" is related here to a coordinate – the Z coordinate; cf. Remark 1 on p.13). This applies to 18 canonical AAs in correspondence to 18 hypothetical "ribosomal" codons (cf. Fig. 2, which depicts the basic spiral of canonical AAs).

(e) Purines are seen as more complex than pyrimidines[5]; a G-C base pair is held together by three hydrogen bonds and provides a greater stabilizing influence than A-U pair, which has only two hydrogen bonds. G in position **Z** leads to the digrams - **.**UA and **.**GU – to inversion, i.e. transformation to complemental rs/codon (cf. last two codons in column "m/codon" in Table 3, including the explanation in legend). With this in view, Figure 2 depicts a hyper-spiral K to **M**. On the other hand, Table 3 summarizes the relations of (numeric) m/codon to rs/codon transformation, as well as mRNA - canonical AAs mapping. (Ordinal numbers are given in decimal numbering system.). In addition it is important to say that Figure 5 suggest two ideas: (1) the canonical AAs spiral creeps under "phantomic digram" •GA **(phd)**, making evident a spiral in space; (2) it evokes analogy with *sensory code*, which is the matter of further researches (cf. last paragraph in Section 3).

## 5 PHYSICO-CHEMICAL PARAMETERS

Physico-chemical parameters of canonical amino acids such as hydrophobicity and polarity, are closely related to the parameters of the above mentioned mapping hyper-spiral. So, Tables 4 and 5 are related to Figure 1, 3 and 4. As it is self-evident from the illustrations and their legends, there is a full balance; the balance between ordinal order of AAs, atom number, as well as nucleon number, within side chains of amino acid sub-classes.

---

[5] Purines are seen as energents too, with a role analogous to that in mitochondrial "electronic respiration"; also their "H-potential" makes Guanine, with 3 H-bridges, in a way dominant.

Between all other evident we reveal some hidden balances. So, the number of atoms within amino acids (their side chains) displaying in two last vertices of logical square in Table 4 ("0" – KN,QH,ED,Y and "3" – I,L,V,F) equals 77+50 = 127, which number is last point in a 7-bit binary-logical tree. On the other hand, within AAs displaying in left-directed complement (**32-31-21-30-20-10**), in Table 5, there are 79 atoms, which number is Golden mean point in a 7-bit binary-logical tree; in right-directed complement (**01-02-03-12-13-23**) there are 11 atoms less, and in non-complement (00-11-22-33)[6] still 11 atoms less, what means a determination by both principles – of minimum change and continuity.

Tables 6 shows the grouping of amino acids accordingly to ordinal number and to two patterns of logical square at the same time. On the other side, in Table 7 it plays only the logical square (in a vice versa position in relation to table 6) but not the ordinal numbers system.

Tables 6 and 7 represent amino acid arrangement based on the order of mRNA *digrams* (Damjanović, 1998); two tables in complemental order, grouped in four logical square patterns (0,1,2,3). The pairs in Table 6, **K**-N, **Q**-H, **E**-D, **Y**-W, **T**-M, **P**-R, **A**-G, **S**-C, **I**-V, **L**-F, with 102 atoms in 10 first pair-members (bold) and 102 atoms in 10 second pair-members, reveal a full symmetrical and proportional balance. There are other balances as well. For example, in three designated amino acid classes there are: 50, 59 and 95 of atoms. The number 59 represents an increase for exactly one modular cycle (in module 9) in relation to number 50; the number 95 is an inversion of 59.

In Table 7 the same pairs appear in a different order: **K**-N, **T**-M, **I**-V, **Q**-H, **P**-R, **L**-F, **E**-D, **A**-G, **Y**-W, **S**-C. Except for the proportional balance 102:102 = 1:1, the system is balanced by number of atoms within amino acid molecules, located on odd and even positions in both lines, the first as well as second pair-members, 61/60 and 41/42, respectively.

These new amino acid pairing agree with physico-chemical properties. Thus, the pairs K-N and Q-H come from a specific crossing: the pair K-H consisting of two basic AAs (the third R is paired with P, both with untypically bonded nitrogen); N-Q as a classic pair – all with nitrogen. Further, from classic H-W, follows Y-W, rather than F-W, because F must go with L, where both molecules possess the same structural motif (isobutane type of branching and H-C-H group between "haed" and "body"). Finelly, both T-M molecules are methyl-derivatives: threonine possesses H-C-CH$_3$ group derived from H-C-H; methionine possesses S-CH$_3$ derived from S-H. The rest of four pairs (E-D, A-G, S-C and I-V) represent the four classical pairs.

---

[6] This classification into two complemental and one non-complemental amino acid classes one must cf. with an analog classification determined by Golden mean (Rakočević, 1998, Scheme 2, p. 289); cf. also App. 2, Survey 3 in this paper.

Table 8 shows a strict distinction in hydrophobicity among canonical AAs in classic AAs pairs (Black and Mould, 1991; Rakočević, 2000). Namely, more hydrophobic (bold) and less hydrophobic (non-bold) canonical AAs appear alternatively, in separate groups, knowing that in the system of classical amino acid pairs (Dlyasin, 1998; Rakočević, 2004), the order of canonical AAs, in correspondence with the hydrophobicity, is as follows from Fig. 5: (**K**-R, **Q**-N, **E**-D, **C**-S, **I**-P, **L**-A, **V**-G, **F**-Y, **W**-H, **M**-T) / (**K**, **Q**, **E**) (T, P, A, S, R, G) (**C**, **I**, **L**, **V**, **F**) (N, H, D, Y) (**W**, **M**). At the same time, the more hydrophobic canonical AAs (first members) are handled by class I enzymes, aminoacyl-tRNA synthetases (all but lysine and phenylalanine; lysine as a "pure" amino derivative and phenylalanine as a "pure" aromatic amino acid), whereas less hydrophobic (second members) canonical AAs are handled by class II enzymes (all but arginine and tyrosine). It is also self-evident, from Table 8, that these regularities are followed by a balance of the number of atoms within amino acid side chains (numbers for the brackets) and of the sums of ordinal numbers (numbers within the brackets) (cf. legend of Table 8).

The splitting into AAs groups after hydrophobicity corresponds with the splitting after polarity (Figure 5); after polar requirement, cloister energy and hydropathy (Kyte and Doolittle, 1982; Rakočević, 2004). Canonical AAs on the left side of horizontal cross leg, in Fig. 5, are polar (all but alanine), whereas canonical AAs on the right side are nonpolar (all but arginine). On the other hand, at the vertical cross leg only two outer canonical AAs (M and F) are nonpolar, whereas all other – the inner canonical Aas – are polar. (Certainly, one must bear in mind that G and P are ambivalent).

## 6  CONCLUSION

A new understanding of codon – amino acid assignment, displayed through previous five Sections, appears – through presented regularities – to be very adequate from the aspect of core essence of coding process within genetic code. Between all others aspects, it is showed that positions of 20 canonical amino acids and belonging codons are arranged through a specific – spiral as well as a cross model of Genetic Code, such a model which stay in correspondence with physico-chemical properties of canonical amino acids and with the sequence of natural numbers at the same time.

# APPENDICES

|     | I     | II    | III     | IV      | V    |           |           |
|-----|-------|-------|---------|---------|------|-----------|-----------|
| IV  | 03 ⊗  | 04 T  | 11 C,W, ⊗ | 12 I,M,I | 19 Y | **83** (518) | **126+10** |
| III | 02 E  | 05 P  | 10 G    | 13 L    | 18 D | **39** (231) | (750 − 1) |
| II  | 01 Q  | 06 A  | 09 R    | 14 V    | 17 H | **53** (311) | **126-10** |
| I   | 00 K  | 07 S  | 08 S,R  | 15 F,L  | 16 N | **77** (440) | (750 + 1) |
|     | 36    | 25    | 63      | 87      | 41   |           |           |
|     |       |       | (88)    | (87)    |      |           |           |
|     |       |       |    (77) |         |      |           |           |

**App. 1, Survey 1**. Amino acid order read from GCT (I). The splitting (classification and/or grouping) of canonical amino acids in accordance to their ordinal number, given by first variant of spiral model of genetic code, as it is presented in Remark 1 on p. 13. The ordinal numbers are reading from codons and/or their digrams (doublets) in the quaternary numbering system. A strict balance in atom number and nucleon number is self-evident. The three AAs groups (77-78-88), read from columns, stay in correspondence with Golden mean balance (60-66-78) as it is shown in App. 2, Survey 3. The reading from the rows gives an atom number balance: 83+53 = 126+10 and 39+77= 126-10; then, a nucleon number balance: 518+231=750-1 and 311+440 = 750+1. (Note: the number of atoms within amino acid molecules as in Rakočević and Jokić, 1996; and nucleon number as in Shcherbak, 1993, 1994).

|     | I      | II        | III   | IV    | V     |           |              |
|-----|--------|-----------|-------|-------|-------|-----------|--------------|
| IV  | 03 ⊗   | 04 T,**W**| 11 C  | 12 I  | 19 Y  | 59 (386)  | 102-1        |
| III | 02 E   | 05 P,**M**| 10 G  | 13 L  | 18 D  | 50 (306)  | (628+10)     |
| II  | 01 Q   | 06 A      | 09 R  | 14 V  | 17 H  | 53 (311)  | 102+1        |
| I   | 00 K   | 07 S      | 08 ∅  | 15 F  | 16 N  | 42 (252)  | (627-10)     |
|     | 36     | 54        | 23    | 50    | 41    |           |              |
|     |        |    (77)   |       | (50)  |       |           |              |
|     |        |           | (77)  |       |       |           |              |

**App. 1, Survey 2**. Amino acid order read from GCT (II). All as in previous Survey except the amino acids are given as one-meaning. A strict balance in atom number and nucleon number is also self-evident. The three AAs groups (77-77-50), read from columns, stay in correspondence with Golden mean balance (60-66-78) as it is shown in App. 2, Survey 3. The reading from the rows gives an atom number balance: 59+42 = 102-1 and 50+53= 102+1; then, a nucleon number balance: 386+252=628+10 and 306+311 = 627-10.

|   |     |       |     |   |     |       |     |
|---|-----|-------|-----|---|-----|-------|-----|
|   | 02  |       |     |   | 06  |       |     |
| 5 | 13  | 120   |     |   | 12  | 162   |     |
|   | 24  |       |     |   | 24  |       |     |
|   | 35  |       |     | 6 | 30  |       |     |
|   | 46  | (084) |     |   | 42  | (42)  |     |
|   | 57  |       | 102 |   | 48  |       |     |
|   | 68  | 204   |     |   | 60  |       |     |
| 3 | 79  |       |     | 3 | 66  | 204   | 402 |
|   | 90  | (222) |     |   | 78  |       |     |
|   | 101 |       |     |   | 84  |       |     |
| 4 | 112 | 426   |     |   | 96  | (444) |     |
|   | 123 |       |     | 6 | 102 |       |     |
|   |     |       |     |   | 114 |       |     |
|   |     |       |     |   | 120 | 648   |     |
|   |     |       |     |   | 132 |       |     |

**App. 2, Survey 1 (left).** Arithmetical regularities as determinants of Genetic code (I). The arithmetical regularities that determine the splitting of amino acids into classes possesses 57-68-79 of atoms (Table 5); a determination through the connection with the total number of atoms (204) within 20 AAs molecules, i.e. their side chains. Notice also a parallel determination through Pythagorean pattern: 3-4-5.

**App. 2, Survey 2 (right).** Arithmetical regularities as determinants of Genetic code (II). The arithmetical regularities that determine the splitting of amino acids into classes possesses 60-66-78 of atoms (cf. App. 2, Survey 3); a determination through the connection with the total number of atoms (204) within 20 AAs molecules (their side chains), and through Golden mean at the same time. Notice also a parallel determination through first perfect number (6) and its half (3).

| 60 | 66 | 78 | 87 | 87 |
|----|----|----|----|----|
| −  | +  | −  | −  | −  |
| 10 | 11 | 01 | 10 | 01 |
| 50 | 77 | 77 | 77 | 88 |

**App. 2, Survey 3**. Arithmetical regularities as determinants of Genetic code (III). The connection of arithmetical regularities presented in two previous Surveys.